\documentclass[pre,onecolumn,assymb]{revtex4}
\usepackage{epsfig,amsfonts,amsmath,bbm}
\usepackage{color}
\definecolor{RED}{rgb}{1.0,0.0,0.0}
\definecolor{BLUE}{rgb}{0.0,0.0,1.0}
\newcommand{\be}{\begin{equation}}
\newcommand{\ee}{\end{equation}}

\begin{document}

\title{Optimal orientation in branched cytoskeletal networks}

\author{D. A. Quint$^1$ and J. M. Schwarz$^1$}
\affiliation{$^1$Physics Department, Syracuse University, Syracuse, NY 13244}

\date{\today}

\begin{abstract}
Actin cytoskeletal protrusions in crawling cells, or lamellipodia, exhibit various morphological properties such as two characteristic peaks in the distribution of filament orientation with respect to the leading edge. To understand these properties, using the dendritic nucleation model as a basis for cytoskeletal restructuring, a kinetic-population model with orientational-dependent branching (birth) and capping (death) is constructed and analyzed. Optimizing for growth yields a relation between the branch angle and filament orientation that explains the two characteristic peaks. The model also exhibits a subdominant population that allows for more accurate modeling of recent measurements of filamentous actin density along the leading edge of lamellipodia in keratocytes. Finally, we explore the relationship between orientational and spatial organization of filamentous actin in lamellipodia and address recent observations of a prevalence of overlapping filaments to branched filaments---a finding that is claimed to be in contradiction with the dendritic nucleation model. 
\end{abstract}

\maketitle
 
\section{Introduction}
The process of cell motility involves a number of components---the actin cytoskeleton, the cellular membrane, an assortment of actin-binding proteins, molecular motors and integrins---that assist the cell in changing shape so that it can move in a particular direction~\cite{Bray}. Naturally, one assumes that the interplay between the various components has been tuned to form structures that optimize for efficient motility. To test this assumption quantitatively is not necessarily an easy task given the dynamically complex structures that emerge as a cell crawls. However, theoretical descriptions of complex biological systems rooted in simplicity may help to identify key interactions so that the sophistication of cell motility may be better quantitatively understood~\cite{TheriotReview,MogilnerReview}.

In order for the cell to crawl in a particular direction, the cell extends itself. This extension, otherwise known as the lamellipodium, is facilitated by the growth and restructuring of the actin cytoskeleton. Over the past ten years or so, the dendritic nucleation model has emerged as the dominant conceptual picture of this reorganization~\cite{Mullins,Pollard}. The dendritic nucleation model asserts that cytoskeletal growth is initiated by membrane-bound proteins, such as PIP$_2$, that activate WASP. WASP, in turn, activates Arp2/3, a protein that nucleates new filaments from preexisting ones. At the point of nucleation, the branch angle takes on a somewhat regular angle of 70 degrees with respect to the mother filament~\cite{Mullins,Svitkina}. This nucleation takes place at/near the cell membrane and leads to a tree-like, or dendritic structuring of the actin cytoskeleton. 
        
New filament growth must be accompanied by some system of regulation, since unregulated birth of branched actin filaments can lead to a redundant use of finite resources in a cell. It has been shown in purified reconstituted systems thatArp2/3 and G-actin alone are insufficient for motility~\cite{Loisel,Carlier}. Additional regulation of the existing actin cytoskeleton is required for rapid G-actin turnover. This regulation is partially assisted by capping proteins, which attach to the plus ends of filaments and stop polymerization. In other words, the filament dies. Also, filaments further back from the leading edge debranch and get severed, eventually becoming part of the finite pool of G-actin. All of these processes are qualitatively described by the dendritic nucleation model. For a unified quantitative description of such processes see Ref.~\cite{Gopinathan}. 

Using the branching (birth) and capping (death) processes as a basis for formation of lamellipodia, we give quantitative evidence to support the notion that form/morphology of the dendritic network is optimized to facilitate cell motility. More specifically, we propose a mean field model for the birth and death rates as a function of filament orientation with respect to the leading edge. We then optimize for filament reproduction at the leading edge, which provides an optimal relation between the branch angle and the angle with respect to the leading edge that agrees with experimental observation~\cite{Svitkina}.

We must point out that there exists an earlier mean field model, which predicts the same optimal relation between the branch angle and the angle of orientation with respect to the leading edge. However, the earlier model has a different physical basis~\cite{Maly}. In keeping with the scientific method, we study further implications of the two models in order to make other retrodictions/predictions to distinguish them. For example, by studying the implications of our model on the spatial organization of filaments, we propose a new shape for the filament density along the leading edge. The shape may account for an observed ``excess'' filament density along the outer edges of the lamellipodia beyond what current modeling predicts~\cite{Keren}. We also make comparisons with a more recent orientational model~\cite{Weichsel}.

Finally, our analysis of spatial information allows us to investigate a recent experimental study of lamellipodia made by Urban and collaborators using electron tomographic images of cytoskeletal networks~\cite{Urban}. They found that overlapping actin filaments were much more prevalent than branched filaments. Based on this data, they proposed an alternate model for the reshaping of actin filaments near the leading edge---that polymerization and cross-linking are the main ingredients for cellular extension and not Arp2/3, which is relegated to a non-branching nucleating agent of new filaments just as dimerization nucleates new filaments. We implement a discrete, spatial simulation of our model to measure, for example, the ratio of overlaps to branch points to determine if the prevalence of overlaps rules out the dendritic nucleation model. We also use the full two-dimensional spatial information of the filament tips to discuss implications for the buckling of the network. 

The paper is organized as follows: Section II introduces and analyzes the mean field, orientational birth-death model. Comparisons with earlier mean field models, as well other generalizations, are addressed. Fluctuations about the mean field solutions are investigated. Section III studies the coupling between the orientational degrees of freedom and the spatial degrees of freedom, such as analyzing the full two-dimensional information of filament positions via discrete simulations.  Section IV discusses the implications of our results. 

\section{Mean field Models}

\subsection{Collision-based model}

Actin filaments contain an inherent polarity where actin monomers associate with the plus end of the filament and dissociate from the minus end~\cite{Moore, Woodrum, Holmes}. This polarity allows for directed assembly such that the cell can extend itself in a particular direction. While extension via polymerization is one mechanism for extension, the dendritic nucleation model~\cite{Mullins,Pollard} asserts that extension via nucleation of branched filaments off pre-existing ones is also important.  Support for the dendritic nucleation model has come about, for example, from electron micrograph images~\cite{Svitkina} of branched actin networks in lamellipodia, from the knocking out of Arp2/3 preventing the formation of lamellipodia~\cite{May} and from reconstituted systems of purified proteins~\cite{Loisel}. In these reconstituted systems, motility can be induced by using a small number of purified proteins combined {\it in vitro} which can reach speeds, for optimal concentrations, of 2.2 $\mu m$ $min^{-1}$. It was observed that motility cannot occur with activated Arp2/3 alone. Additional proteins, which facilitate a \emph{steady state} of G-actin concentration, are essential~\cite{Loisel}. These proteins are capping proteins, which cap polymerizing plus ends, and ADF/Cofilin which cuts actin filaments. Both proteins help replenish the G-actin pool. 

To test some of the dendritic nucleation model assertions, let us construct a mean field model with branching and capping and investigate various experimental consequences. As for the branching, {\it in vitro} studies suggest a preferred angle of $70^{\circ}$ with respect to the plus end of the mother filament~\cite{Mullins}.  Therefore, for now, we assume that the branching angle is some fixed angle $\psi$ from the plus end. We will also assume that Arp2/3 branches off the side of pre-existing filaments with a preference towards the plus end as has been observed experimentally~\cite{Ichetovkin}. Moreover, we assume that the nucleation of a branch occurs at and/or very near the membrane. Of course, if branching takes place only at the membrane, then the initial structure of the network is dictated by the shape of the membrane.  If the Arp2/3 is released from the membrane upon activation and then collides (binds) with actin filaments, then the network structure is less dependent on the shape of the membrane.  Recent experiments observed space-filling polymerization of filopodia into gaps between the edge of the network and the membrane~\cite{Yang}. Such an observation has to yet found with Arp2/3 nucleation, however. 

So, assuming that side-branching occurs and that the branch is nucleated
at/near the membrane, the branching probability depends on the orientation of
the filament. The more the filament is parallel with the leading edge, the
higher the cross-section for collision between the globular Arp2/3 and the
one-dimensional filament and, hence, nucleation.  More precisely, the
branching rate contains a $|\sin(\theta)|$ dependence, where
$\theta=0^{\circ}$ is normal to the leading edge. See Figure 1. 

As for the death rate, filament plus ends get capped at a rate $c$. We will not assume any angular dependence for the capping rate. The capping protein-plus end binding is a globular-to-globular collision. Furthermore, the task of the capping protein is to regulate the length of filaments such that growth is channelled into developing new filaments and not into extending pre-existing ones~\cite{CarlssonCapping}. Elongating pre-existing filaments leads to a system longer filaments on average, which are more susceptible to buckling~\cite{Odijk}. Channeling new filament growth should, therefore, be independent of filament orientation. In addition, channelling growth into branches allow for further spreading of lamellipodia, which increases cell contact with the surface in order to build more focal adhesions. 

\begin{figure}[htb]
\begin{center}
\includegraphics[width=5cm]{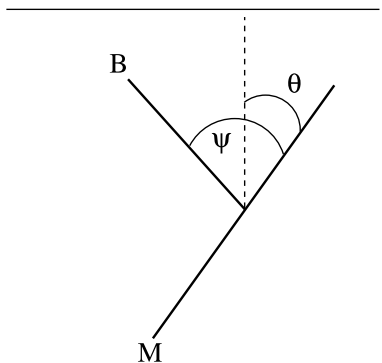}
\caption{Schematic of the orientation of a branched filament (B) in relation
  to its mother filament (M) and the horizontal line denotes the leading
  edge. The dashed line represents the normal to the leading edge.}
\end{center}
\end{figure} 

Based on the above assumptions, we construct a set of kinetic equations that take into account the orientation of filaments, which branch off prexisting filaments and get capped. We restrict ourselves to $-90^{\circ}<\theta<90^{\circ}$ since we are only interested in ``forward'' growth. We first consider $\psi>45^{\circ}$ and $0^{\circ}<\theta<\psi$. In this regime, there are are two populations of filaments, filaments oriented at angle $\theta$, denoted by $n_1$, and filaments oriented at an angle $\theta-\psi$, denoted by $n_2$. (There is a reflection symmetry about $\theta=0^{\circ}$. We will only deal with $0^{\circ}<\theta<90^{\circ}$ and use the reflection symmetry to extend our results to $-90^{\circ}<\theta<0^{\circ}$.) The kinetic equations for this first case are

\begin{equation}
\frac{dn_1}{dt}=\frac{b}{2}|\sin(\theta-\psi)|n_2-cn_1
\end{equation}
and
\begin{equation}
\frac{dn_2}{dt}=\frac{b}{2}|\sin(\theta)|n_1-cn_2,
\end{equation}
where $b$ denotes the magnitude of the branching rate. The factor of 1/2 is because branching on the ``backside'' of the mother filament is a less-likely collision given the activation of Arp2/3 at the membrane and we do not consider it here. 

We now rephrase famous ``the form follows function'' optimization guideline as a population biology problem~\cite{Bock,Fisher,Crow}. We assume the cytoskeletal system is maximizing for ``population'' growth so that the cell can extend itself efficiently. To determine this maximal growth, we compute the eigenvalues for the above set of equations and determine the relationship between $\theta$ and $\psi$ such that the largest of the two eigenvalues is maximized (and positive). The eigenvalues for the above set of equations are

\begin{equation}
\lambda_{1,2}=-c\pm \frac{b}{2}\sqrt{|\sin(\theta)||\sin(\theta-\psi)|}.
\end{equation}

It is easy to see that the largest eigenvalue is maximized when $\theta^*=\psi/2$.  Of course, $\theta^*=-\psi/2$ is another optimal solution via symmetry. 
\\
 
Next, we investigate $\psi<\theta<90^{\circ}$ (and $\psi>45^{\circ}$). In this second regime, there are three orientations of filaments with $n_3$ denoting filaments oriented at $\theta-2\psi$. The set of kinetic equations for this second case are

\begin{eqnarray}
\frac{dn_1}{dt}&=&\frac{b}{2}|\sin(\theta-\psi)|n_2-cn_1,\\
\vspace{4pt}
\frac{dn_2}{dt}&=&\frac{b}{2}|\sin(\theta)|n_1+\frac{b}{2}|\sin(\theta-2\psi)|n_3-cn_2,\\
\vspace{4pt}
\frac{dn_3}{dt}&=&\frac{b}{2}|\sin(\theta-\psi)|n_2-cn_3.
\end{eqnarray}

Once again, assuming activation of Arp2/3 at the membrane, nucleation is unlikely to occur on the ``backside'' of a mother filament with respect to the membrane and the kinetic equations become

\begin{eqnarray}
\frac{dn_1}{dt}&=&-cn_1,\\
\vspace{4pt}
\frac{dn_2}{dt}&=&\frac{b}{2}|\sin(\theta)|n_1+\frac{b}{2}|\sin(\theta-2\psi)|n_3-cn_2,\\
\vspace{4pt}
\frac{dn_3}{dt}&=&\frac{b}{2}|\sin(\theta-\psi)|n_2-cn_3.
\end{eqnarray}

The $n_1$ population eventually dies off such the above equations simplify further to 

\begin{eqnarray}
\frac{dn_2}{dt}&=&\frac{b}{2}|\sin(\theta-2\psi)|n_3-cn_2,\\
\vspace{4pt}
\frac{dn_3}{dt}&=&\frac{b}{2}|\sin(\theta-\psi)|n_2-cn_3.
\end{eqnarray}

With the transformation of $\theta'=\theta-\psi$, we map back to the first set of kinetic equations such that the largest, positive eigenvalue occurs again at $\theta'^*=\pm\psi/2$. If $\psi=45^{\circ}$, there exists another optimum at $\theta^*=67.5^{\circ}$. However, this initial orientation will die away and the $\pm 22.5^{\circ}$ will survive. So we have the same optimization as in the first case, i.e. it is redundant. This result is different from the initial model~\cite{Maly}, where the optima occur at $\theta=0^{\circ}, \pm \psi$.  Of course, for $\theta=0^{\circ}$, the critical buckling load is the smallest, i.e. filaments are more susceptible to buckling. While this property is not optimal for rheology, bundled filaments can increase the critical buckling load~\cite{Rubinstein}. For our model, there can be no optimum at $\theta=0^{\circ}$. 

If we increase $\psi$ beyond $60^{\circ}$, the second optimization peak is outside of the range of interest ($\psi<\theta<90^{\circ}$). However, consider $\psi=70^{\circ}$. As $\theta$ increases from $70^{\circ}$ to $90^{\circ}$, the reproductive growth enhances monotonically. For $\theta=90^{\circ}$, the daughter filaments are subsequently oriented at $20^{\circ}$ and $-50^{\circ}$.  While these initial $90^{\circ}$ filaments are not precisely optimized for growth, their growth is maximum for the interval, $\psi<\theta<90^{\circ}$. Therefore, we deem these $20^{\circ}$ and $-50^{\circ}$ filaments as suboptimal orientations. Figure 2 depicts the optimal orientations and the suboptimal, or subdominant, orientations. We conjecture that the subdominant orientations of filaments may serve as reinforcements for cross-linking.  Depending on the initial spatial arrangement and orientation of the filaments, the subdominant orientations may help to increase overlaps and spreading. It is interesting to note that only for $\psi>60^{\circ}$, the second redundant optimum is removed. The observed branch angle is reasonably close to this value. 
\begin{figure}
\begin{center}
\includegraphics[width=4cm]{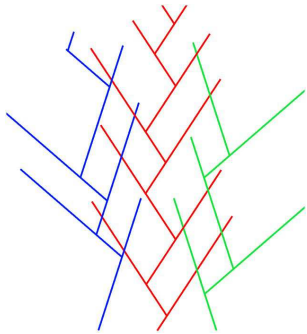}
\caption{Depiction of the optimal orientation (red) and the two suboptimal ones (blue, green).}
\end{center}
\end{figure} 

Experiments on keratocytes have measured the distribution of orientation of filaments normal to the leading edge~\cite{Maly}. There are two maxima in the distribution occurring at $\pm 35^{\circ}$. Assuming that the branch angle is indeed $70^{\circ}$, our optimization analysis provides an explanation for this experimental finding. However, we should mention that Koestler and collaborators have conducted a more recent experiment on the orientation of filaments~\cite{Koestler}. They observed a broad distribution between the angles of $-75^{\circ}$ to $75^{\circ}$. One could argue that the subdominant orientation of filaments could account for further spreading of the distribution and, therefore, perhaps the more recent data.

\subsection{Comparison with other models}

How does the above model compare with the one constructed and analyzed by Maly and Borisy~\cite{Maly}? The Maly/Borisy model is consistent with the Brownian ratchet model~\cite{Peskin,MogilnerOster} for filament elongation near a membrane. In the Brownian ratchet model, leading edge filaments polymerize only if there exists enough space between the membrane and the tip of the filament. As the filaments fluctuate, transient gaps open up between the filament and the membrane, allowing actin monomers to attach to the plus end of the filament. Once the filament bends back to its original straight configuration, it is now longer and, therefore, pushes against the membrane moving it forward. This process, however, is limited to the size of the fluctuations that occur between the membrane and the tip of the leading edge filament. In support of this notion, experiments involving changing the membrane tension have shown that there exists an inverse relationship between the lamellipodial extention velocity and the apparent membrane tension~\cite{Sheetz}. However, more recent experiments suggest more complicated mechanisms may be at play~\cite{Prass}.

Given the space limitation between the membrane and the fluctuating tip, there is an orientational degree of freedom that the filaments can exploit in this polymerization process. By varying the angle at which the tip makes with the membrane initially, the amount of space between the two can change if the membrane moves forward over a time $t$ at a velocity $v_{mem}$. In particular, $\delta k_p \rho_m p \cos(\theta)= v_{mem}$, where $\delta$ is the G-actin diameter, $k_p$ is the polymerization rate, $\rho_m$ is the G-actin concentration, and $p$ is the probability that the filament tip is not obstructed by the membrane. Maly and Borisy~\cite{Maly} assert that the capping of a filament is only possible if the growing filament tip is not obstructed by the membrane, hence capping occurs at a rate $cp$. Since $p\propto 1/\cos(\theta)$, the larger $\theta$ is, the more likely the filament will be capped. 

As for the branching, in the Maly/Borisy model, the branching rate does not contain any angular dependence.  In other words, the kinetic equations read  

\begin{eqnarray}
\frac{dn_1}{dt}&=&\frac{b}{2}n_2-\frac{cp_0}{\cos(\theta)}n_1\\
\frac{dn_2}{dt}&=&\frac{b}{2}n_1-\frac{cp_0}{\cos(\theta-\psi)}cn_2,
\end{eqnarray}
where $p_0=v_{mem}/\delta \rho_m k_p$. Maly and Borisy show that for $\cos(\psi)<p_0<\cos(\psi/2)$, the optimal relation between $\theta$ and $\psi$ is, as above, $\theta^*=\pm \psi/2$. However, for $p_0<\cos(\psi)$, the optimal orientations are zero and $\pm \psi$.  

A more recent orientational model assumes a $\theta$-independent, zeroth-order branching rate, a $\theta$-independent, first-order capping rate, and a $\theta$-dependent outgrowth rate that kills single filaments outgrowing the bulk of the network~\cite{Weichsel}. The model exhibits two different, stable patterns, the same two exhibited by Maly and Borisy~\cite{Maly}, $\theta=\pm\psi/2$ or $\theta=0,\pm\psi$. The two patterns cannot coexist. Parameters such as the capping rate determine which pattern prevails.  The authors argued that their model can explain the experimentally observed load-dependence of the network velocity at a given force~\cite{Parekh}. Our model cannot exhibit the latter pattern and our subdominant pattern for $\psi>60^{\circ}$ does coexist with the primary, or optimal, one.    

\subsection{Generalized Birth/Death rates}

While each mean field model has a different physical basis, the selection criterion for maximal growth yields the same optimal relationship, $\theta^*=\pm\psi/2$. How generic is this result? To begin to answer this, we consider the most general version of our population equations such that both the birth-rate and the death-rate depend on the orientation of the branched filaments. Therefore, we begin with
\begin{eqnarray}
\frac{dn_1}{dt}&=&B_{1}(\theta,\psi)n_2+D_{1}(\theta,\psi)n_1\\
\frac{dn_2}{dt}&=&B_{2}(\theta,\psi)n_1+D_{2}(\theta,\psi)n_2,
\end{eqnarray}
for $0^{\circ}<\theta<90^{\circ}$. We define the matrix, $\mathbf{Q}\equiv\mathbf{Q(\theta,\psi)}$ such that we can represent the set of linear coupled equations vectorially as $\mathbf{\dot{n}}=\mathbf{Q}\mathbf{n}$, where 
\begin{center}
\begin{equation}
\textbf{Q}=
\begin{pmatrix}
D_{1}(\theta,\psi) & B_{1}(\theta,\psi) \\
B_{2}(\theta,\psi) & D_{2}(\theta,\psi) \\
\end{pmatrix}.
\end{equation}
\end{center}

Defining $\mathbf{\overline{Q}}=\mathbf{Q}/Det[\mathbf{Q}]$, the eigenvalues of $\mathbf{\overline{Q}}$ are given by 

\begin{equation}
\lambda_{+,-} = \frac{Tr[\mathbf{\overline{Q}}]\pm\sqrt{Tr[\mathbf{\overline{Q}}]^{2}-4}}{2}.
\end{equation}
With this result, three scenarios emerge:
\begin{equation}
\begin{array}{cc}
Condition & Eigenvalues\\
(1)\hspace{5 pt} |Tr[\mathbf{\overline{Q}}]| < 2  \hspace{5 pt} &\lambda_{+,-}\rightarrow \mathbbm{C} \\
&\\ 
(2)\hspace{5 pt} |Tr[\mathbf{\overline{Q}}]| = 2   \hspace{5 pt} &(\lambda_{+} = \lambda_{-})\rightarrow \mathbbm{R}\\
&\\
(3)\hspace{5 pt} |Tr[\mathbf{\overline{Q}}]| > 2 \hspace{5 pt} &(\lambda_{+}>\lambda_{-})\rightarrow \mathbbm{R}
\end{array} 
\end{equation}

To determine the largest, real eigenvalue, we focus on condition 3. Dropping the +,- notation, the optimization condition is determined by 
\begin{equation}
\partial _\theta\lambda= Tr[\partial\mathbf{\overline{Q}}]\bigg(1+\frac{Tr[\overline{Q}]}{\sqrt{Tr[\mathbf{\overline{Q}}]^{2}-4}}\bigg)=0
\end{equation}
such that
\begin{equation}
Tr[\partial_{\theta} \mathbf{\overline{Q}}]=0.
\end{equation}
In other words, the optimization condition occurs when the matrix $\partial_{\theta} \mathbf{\overline{Q}}$ is rendered traceless. One, of course, also needs to evaluate the second derivative to check for a maximum. 

With the help of the Jacobi formula for the derivative of the determinant of a matrix and using the linearity of the trace operator, the optimization condition for $\mathbf{Q}$ must satisfy (assuming the trace of $\partial_\theta \mathbf{Q}$ is zero),
\begin{equation}
Tr[\mathbf{Q}^{-1} \partial_{\theta} \mathbf{Q}]=0.
\end{equation}
If we analyze the case where the two death rates are $\theta$-independent, then the optimal condition is 

\begin{equation}
\partial_{\theta}(B_{1}(\theta,\psi)B_{2}(\theta,\psi))=0.
\end{equation}
For example, if $B_1(\theta,\psi)$ is $\psi$-independent and $B_2(\theta,\psi)=B_1(\theta-\psi)$, then the optimal condition is $\theta^*=\psi/2$ as long as $B_1(\theta)$ is an even function (provided the second derivative is negative at that point). If $B_1$ is a trigonometric function, then the periodicity should not be too small such that other maxima appear within the $0^{\circ}<\theta<90^{\circ}$ range. So, $B_1=\cos(\theta)$ yields the same optimal relation between $\theta$ and $\psi$ as would many other functions. It is possible to broaden this analysis. We leave this for future work. Our point now is that the optimal finding of $\theta^*=\pm\psi/2$ alone does not necessarily justify the model. One needs to explore further implications of the model in order to distinguish it from other potential models.  We shall pursue this tact in the next section.

\subsection{Fluctuations}

Is the optimal relationship between $\theta$ and $\psi$ robust in the presence of fluctuations? To answer this question, following Maly and Borisy~\cite{Maly}, we assume that the angle between the two types of filaments exhibits Gaussian fluctuations with a mean of $\psi$ and a variance of $\sigma^2$. If we define $n(\theta,t)$ as the density of filaments at the leading edge at time $t$ and orientation $\theta$, then the dynamic equation for $n(\theta,t)$ (for $-\psi<\theta<\psi$) is given by
\begin{equation}
\frac{\partial n(\theta,t)}{\partial t}=\frac{b|\sin(\theta)|}{\sqrt{8\pi}\sigma}\int^{\psi}_{-\psi} (e^{-\frac{(\theta'+\psi-\theta)^2}{2\sigma^2}}+ e^{-\frac{(\theta'-\psi-\theta)^2}{2\sigma^2}})n(\theta',t)d\theta'-cn(\theta,t).
\end{equation}
If we assume $n(\theta,t)=N(t)q(\theta)$, we arrive at 
\begin{equation}
\frac{|\sin(\theta)|}{\sqrt{8\pi}\sigma} \int^{\psi}_{-\psi} (e^{-\frac{(\theta'+\psi-\theta)^2}{2\sigma^2}}+ e^{-\frac{(\theta'-\psi-\theta)^2}{2\sigma^2}})q(\theta')d\theta'=\frac{c'}{b'}q(\theta),
\end{equation}
where $\frac{c'}{b'}$ is now an unknown eigenvalue such that the above assumption is justified. We use the quadrature method to numerically solve for $q(\theta)$ for different values of $\sigma$.  Figure 3 depicts the results. The maximum of $q(\theta)$ correspond well with the largest, positive eigenvalue found previously. As $\sigma$ increases, the maxima remain robust, but are become less pronounced. These results indicate that the optimal relation of $\theta^*=\pm \psi/2$ is robust to fluctuations.  
  
\begin{figure}
\begin{center}
\includegraphics[width=9cm]{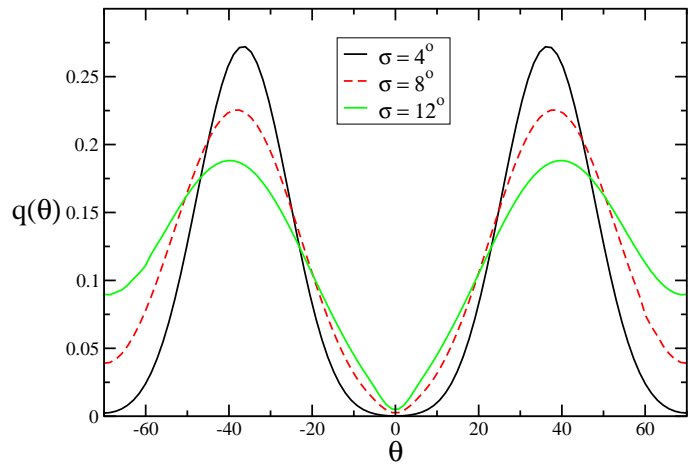} 
 \caption{Plot of $q(\theta)$ for $\psi=70^{\circ}$. For $\sigma=4^{\circ}, \frac{c'}{b'}=0.269$, for $\sigma=8^{\circ}, \frac{c'}{b'}=0.254$, and for  $\sigma=12^{\circ}, \frac{c'}{b'}=0.213$.}
\end{center}
\end{figure}

It is certainly worth comparing this result with the fluctuation results of Maly and Borisy~\cite{Maly}. The steady state orientation in the presence of noise here is very similar to the Maly/Borisy model~\cite{Maly}, at least for intermediate values of $p_0$. This new computational result is, therefore, somewhat nontrivial given that one would assume the fluctuations to be sensitive to the details of the underlying kinetics. Further investigation along the lines of Section IIc is needed to pursue understanding of this possible genericity despite differing details of the kinetics. 

\section{Orientation influencing spatial organization}

\subsection{Filament density profile along the leading edge}

Optimization for growth in lamellipodia leads to a relationship between the branch angle $\psi$ and the orientation of filaments relative to the leading edge, or  $\theta$. To date, there exist three models, each rooted their in own physical basis, that yield $\theta^*=\pm\psi/2$. In order to further differentiate between these models, we investigate the distribution of filament tips along the leading edge.

Previous work investigating the filament density along the leading edge has invoked the following set of assumptions~\cite{Grimm,Lacayo,Keren}. Filaments are either oriented with $+35^{\circ}$ or $-35^{\circ}$ with respect to the leading edge. Their respective densities along the leading edge $x$ are denoted by $\rho^+(x,t)$ and $\rho^-(x,t)$. These filaments undergo lateral flow in their respective directions. Filaments with either orientation can spawn filaments with the opposite orientation ($\pm\rightarrow\mp$) from their own. Also, both types of filaments can get capped. Therefore, the equations for both filament densities along the leading edge, whose position is denoted by $x$, are

\begin{equation}
\frac{\partial \rho^{\pm}}{\partial t} = \mp \frac{\partial}{\partial x}(v\rho^{\pm})+\frac{b}{B} \rho^{\mp}-c\rho^{\pm}
\end{equation}
with $B=\int_{-\frac{L}{2}}^{\frac{L}{2}} dx (\rho^+(x)+\rho^-(x))$, where $L$ is the length of the lamellipodium and $v$ is the lateral flow speed, which is proportional to the speed of the crawling cell.
 
Previous analysis of the above equation yields a total filament density in steady state that is peaked at the center of the cell, provided the filament density at the edges is sufficiently small. More specifically, for the boundary conditions,
$\rho^+(-L/2,t)=0$ and $\rho^-(L/2,t)=0$, $\rho_+(x,t\rightarrow \infty)+\rho_-(x,t\rightarrow \infty)=
\frac{\pi}{2}\frac{b}{Lc}\cos(\frac{\pi x}{L})$. Therefore, near the center the profile is an inverted parabola~\cite{Lacayo,Keren}. If the boundary condition is adjusted to a higher concentration, eventually, the inverted parabola becomes a parabola with the total filament density higher at the sides than in the center.  Assuming $\frac{v}{L}<<c$, we define a dimensionless time, $\tau=ct$, a dimensionless position $s=x/L$, and dimensionless densities, $\tilde{\rho}^{\pm}=\rho^{\pm}\frac{cL}{b}$, then Eq. 25 becomes
\begin{equation}
\frac{\partial \tilde{\rho}^{\pm}}{\partial \tau} = \mp \frac{\partial}{\partial s}(\frac{v}{cL}\tilde{\rho}^{\pm})+\frac{1}{\tilde{B}} \tilde{\rho}^{\mp}-\tilde{\rho}^{\pm}.
\end{equation}
If the boundary conditions at the ends of the lateral extent demand a large enough density, then the system will not be able to sustain the peak in density at the center of the leading edge.  The larger the branching rate, the higher the allowed density at the ends can be with the system still sustaining an inverted parabola.

The inverted parabola in filament density along the leading edge is observed in experiments near its center~\cite{Keren}. However, there is an excess of the filament density towards the sides of the leading edge ($-L/2$ and $L/2$) that appears to be flat. This excess has not been accounted for in the current model. In light of the collision-based model introduced here, we propose that for $\psi<\theta<90^{\circ}$ with $\psi\approx 70^{\circ}$, the subdominant pattern, whose growth is prevented from being fully optimized so as not to form a redundant pattern, may account for this excess. The axis of the subdominant pattern is at $\theta=-15^{\circ}$ as opposed to $\theta=0^{\circ}$. There exists another pair centered at $\theta=15^{\circ}$ by symmetry.  

To test this proposal, we take the simplest approach by constructing the following six equations taking into account the two center populations (as before) and the two respective pairs of subdominant, or ``off-center'', populations. For now, each respective pair of populations is not coupled to any other respective pair. Each pair occupies it own region along the lateral extent of the lamellipodia. We denote $\rho_c^{\pm}$ as the original set, $\rho_l^{\pm}$ as those directed toward the left side of the leading edge (from the birdseye perspective of cell), and $\rho_r^{\pm}$ as those directed toward the right side of the leading edge to arrive at

\begin{eqnarray}
\frac{\partial \rho_c^{\pm}}{\partial t} &=& \mp \frac{\partial}{\partial x}(v \rho_c^{\pm})+\frac{b}{B_c}\rho_c^{\mp}-c \rho_c^{\pm}\\
\frac{\partial \rho_l^{\pm}}{\partial t} &=& \mp \frac{\partial}{\partial x}(v\rho_l^{\pm})+\frac{b_2}{B_l}\rho_l^{\mp}-c \rho_l^{\pm}\\
\frac{\partial \rho_r^{\pm}}{\partial t} &=& \mp\frac{\partial}{\partial x}(v\rho_r^{\pm})+\frac{b_2}{B_r}\rho_r^{\mp}-c \rho_r^{\pm}
\end{eqnarray}
where $B_c=\int_{-\frac{L}{4}}^{\frac{L}{4}} dx (\rho_c^+(x)+\rho_c^-(x))$, $B_l=\int_{-\frac{L}{2}}^{0} dx (\rho_l^+(x)+\rho_l^-(x))$, and  $B_r=\int_{0}^{\frac{L}{2}} dx (\rho_l^+(x)+\rho_l^-(x))$. Note the different spatial regions for each respective pair of populations. Of course, the delineation is not so clear cut in practice. Also, $b_2<b$ to allow for a slight decrease in branching at the edges of the leading edge.  Finally, we assume $v$ does not vary between the different pairs.
 
To solve for the steady state filament density distribution, we use the following boundary conditions. We set $\rho_c^+(-L/4,t)=0$, $\rho_c^-(L/4,t)=0$, $\rho_r^+(0,t)=0$, $\rho_r^-(L/2,t)=\rho_0$, $\rho_l^-(0,t)=0$, $\rho_l^+(-L/2,t)=\rho_0$. As for the asymmetric boundary conditions on $\rho^{\pm}_{r}$ and $\rho^{\pm}_l$, it is reasonable to assume that near the lateral center of the leading edge, the density of $\rho_r^+(0,t)=0$. However, towards the sides of the leading edge, $\rho_r^-(L/2,t)$ may not necessarily vanish as there may be some skewing of the axis along which the subdominant populations are propagating due to the focal adhesions. The same goes for $\rho^+_l$.  For the symmetric boundary conditions for $\rho_c^\pm$, the same cosine steady state solution exists as before (only over a smaller interval).  For the $\rho_r^{\pm}$ and $\rho_l^{\pm}$ populations, the steady state solutions are sinusoidal. (If $\rho_r^+(0,t)=\rho_1<<1$, then the steady state solution is a linear combination of sine and cosine). These solutions are plotted in Figure 4. We use $\frac{v}{Lc}=0.01$, $b_2=0.6b$, and $\tilde{\rho}_0=0.3$. Typical lateral speeds are of order 0.1 microns/sec for fast-moving keratocytes, typical lengths of leading edges are tens of microns and typical capping rates are tenths per second to per second~\cite{Grimm,Pollard}. We have also checked these solutions numerically.

When we sum up the various densities to arrive at the total density, we can account the observed excess filament density at the lateral edges of the lamellipodium while still having the overall proper shape found experimentally near the center of the system. Indeed, there is a small dip in the center of the system, this dip may be difficult to observe experimentally and is presumably washed out once noise and other details are incorporated into the modeling.  For instance, the revised model may be further updated to included coupling between the different populations via branching in terms of the overlapping regions. We leave this for future work. We should also note that the two previous orientational models, at least for $\theta^*=0^{\circ},\pm\psi$,  would yield a filament density distribution that is sensitive to the initial distribution of filaments along the leading edge since the $\theta=0^{\circ}$ does not flow laterally.  Such a sensitivity should be investigated in order to rule out the possibility of the $\theta=0^{\circ},\pm\psi$ pattern.

\begin{figure}
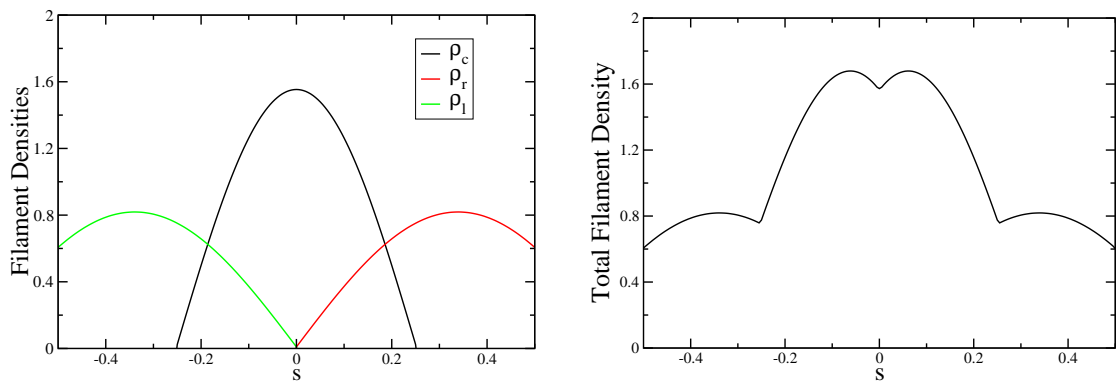

\begin{center}
\includegraphics[width=7cm]{filamentdensities.eps}
\hspace{0.5cm}
\includegraphics[width=7cm]{totalfilamentdensity.eps}
\caption{Left: Dimensionless filament densities along the leading edge, where $\rho_c=\tilde{\rho}_c^+ + \tilde{\rho}_c^-$, for example. Right: Total dimensionless filament density ($\rho_c+\rho_l+\rho_r$) along leading edge.}
\end{center}
\end{figure}

\subsection{Two-dimensional, discrete simulation} 
\begin{figure}[b]
\begin{center}
\includegraphics[width=7cm]{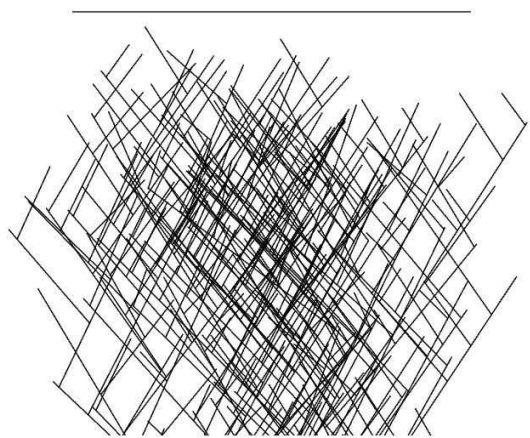}
\caption{Discrete simulation output with $\psi=70^{\circ}$, $\theta=35^{\circ}$, and $\sigma=10^{\circ}$. The length of the horizontal bar is 1 micron.}
\end{center}
\end{figure} 
To further study how filament orientation affects the spatial distribution of filaments, we construct a two-dimensional kinetic simulation with explicit filaments. A two-dimensional approach is reasonable given that lamellipodia are typically flat structures with a thickness of approximately 100 nanometers, extending several microns into the body of the cell and approximately 10 microns across the cell.  The simulation algorithm is as follows:\\
\vspace{0.25cm}

\noindent(0) Initialization: A filament is initialized at the origin of the system with an angle $\theta$ and a length 100 nm.\\
(1) Branching: A random number, $r$, is chosen from the sine distribution. Should $r<\sin(\theta)$ (with $\theta<90^{\circ}$), a branch point is chosen along the initial filament. Where the branch point occurs is uniformly chosen over some part of the current filament length as measured from the growing end. The length is denoted by $f$.  This constraint restricts the branching to occur near the leading edge of the network. The branch filament emerges at an angle $\psi$ with respect to the mother filament. Gaussian fluctuations about the branch angle, with variance $\sigma^2$, are also studied.\\
(3) Capping: A random number, $s$, is chosen uniformly between zero and unity. If $s<c$, the filament gets permanently capped and no longer extends. Also, no further branching can occur along it.\\
(4) Every uncapped filament grows by an additional 100 nm in its initially chosen direction (polymerization).\\
(5) Steps (1)-(4) are repeated for each uncapped filament until capped.\\
\vspace{0.25cm}

Figure 5 demonstrates output from the simulation for a branching angle of 70 degrees with $\sigma=10^{\circ}$. Note that we do not explicitly incorporate a membrane into the simulation and we allow for overlaps of filaments due to the thin, third dimension.  Also, unless specified otherwise, the time step in the simulation is $0.30$ seconds, assuming a constant G-actin concentration of 10 $\mu\,M$, the branch rate is $33.33\,\,s^{-1} \mu m^{-1}$ and the capping rate is $0.83\,\,s^{-1}$~\cite{Grimm,Pollard}.

{\bf Growth:} We first investigate the optimal relation between $\theta$ and $\psi$. In keeping with the mean field analysis, we compute the average number of uncapped filaments generated each time step, denoted as $G$, with an upper bound of 1000 filaments. Note that here we do not distinguish between the two populations, mother and daughter. If the average number of uncapped filaments grows with time, the growth is exponential. Of course, eventually, the system reaches a steady state presumably due to a finite amount of Arp2/3 or other mechanisms.  To study the approach to steady state, one must incorporate recycling of G-actin monomers via depolymerization, severing and debranching as well.  Such mechanisms have been explored by the autocatalytic model developed by Carlsson~\cite{Auto}. In the autocatalytic model, there is an initial overshoot in the number of filaments that has now been observed experimentally~\cite{Kang}.

We present measurements of $G$, averaged over 4000 samples, as a function of $\theta$. Unless otherwise specified, $f=25$ nanometers. We observe in Figure 6 that the optimal relation, $\theta^*=\pm \psi/2$, holds in the two-dimensional simulations. We also see evidence of the subdominant population of filaments for $\theta>\psi$. For $\psi=70^{\circ}$, the growth at $\theta=90^{\circ}$ is more comparable to the growth at $\theta^*$ than for $\psi=80^{\circ}$.  Again, perhaps the subdominant pattern for $\theta=90^{\circ}$ contributes to increased spreading and/or overlaps between generations following the initial filaments. For a smaller capping rate, more growth occurs as
evidenced in Figure 6, though the same optimal relation holds, as expected. When fluctuations are added to the branch angles, the distinguishing feature of zero growth at $\theta^*=\psi$ remains robust as expected. Moreover, $G$ broadens near the maximum. See Figure 7. Broadening was also observed in the mean field simulations with noise.
 
\begin{figure}
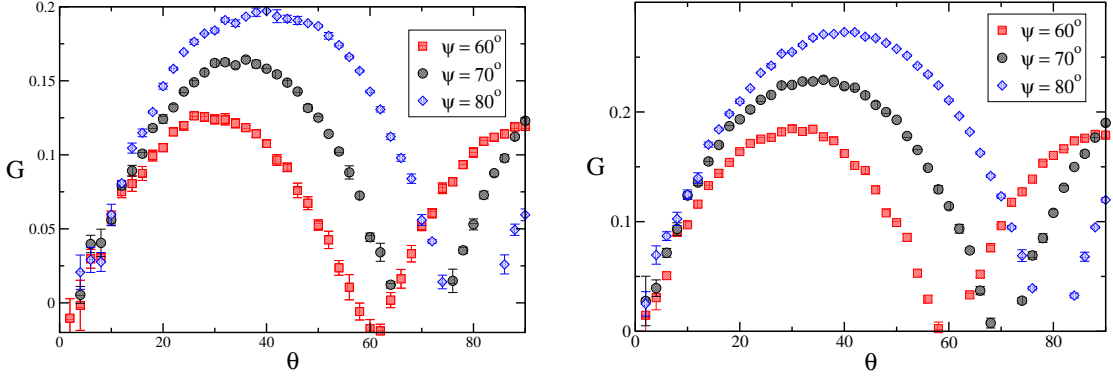

\begin{center}
\includegraphics[width=7cm]{growth.ba.eps}
\hspace{0.5cm}
\includegraphics[width=7cm]{growth.ba.smallercap.eps}
\caption{Left: Growth rate for different branch angles. Right: Growth rate for for $c=0.67\,\,s^{-1}$ (and so more growth than for $c=0.83\,\,s^{-1}$).}
\end{center}
\end{figure} 

\begin{figure}[b]
\begin{center}
\vspace{0.25cm}
\includegraphics[width=7cm]{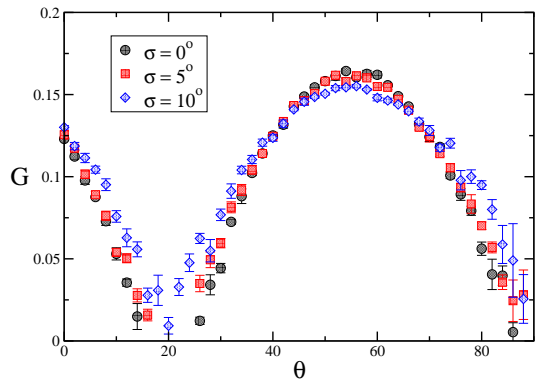}
\caption{Growth rate for 70 degree branching angle with noise.}
\end{center}
\end{figure}

{\bf Overlaps:} Urban and collaborators use electron tomography to observe many more overlaps than branches in lamellipodia~\cite{Urban}. Their technique allows one to probe the three-dimensional aspect of the cytoskeleton such that filaments that appeared to be branches in two-dimensional electron micrograph images turn out to be overlapping
filaments.  Based on the prevalence of overlaps, Urban and collaborators proposed a new model for the structuring of lamellipodia~\cite{Urban}. Arp2/3 nucleates new filaments near the membrane (just as dimerization does) such that there is no pre-existing filament and, hence, no memory of its orientation.  In other words, there is no branching. 

However, we would like to point out that the prevalence of overlaps does not rule out a branched model. In fact, the existence of overlaps is rather natural in a branched model. If each subsequent generation of branches becomes exponentially smaller in length, then there will be no overlaps. This is how one embeds a Bethe lattice--a tree graph---in a plane such that there are no overlaps. See Figure 8. If this exponential decrease in length with each generation does not occur, then overlaps are expected. While the original electron micrographs indicate that the filament length increases
further back towards the cell body, we do not expect an exponential increase. Therefore, there will be crossings/overlaps between branches. In fact, the overlaps can be reinforced by cross-links thereby increasing the temporary rigidity of the network. 
\begin{figure}[t]
\begin{center}
\includegraphics[width=4cm]{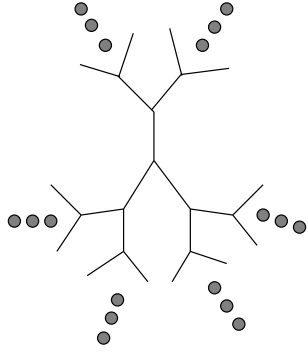}
\caption{Bethe lattice with coordination number three.  The circles indicate a repeating pattern.}
\end{center}
\end{figure} 
To test this idea, we measure the number of overlaps and compute the ratio, $\chi$, of the number of overlaps to the number of branch points for each particular $\theta$. See Figure 9. We see that the ratio peaks where there is optimal growth. Moreover, as the capping rate decreases, the filaments grow longer also allowing for more overlaps. We compare the branched model with a fixed branch angle (plus small fluctuations) to a branched model where the branch angle is uniformly random between 1 and 89 degrees. We do this to disrupt the inheritance in orientation of the fixed branch angle. So, Arp2/3 merely nucleates a filament off a pre-existing filament with no memory of filament orientation.  We model the lack of inheritance with the completely random branch angle to capture some aspect of the recently proposed unbranched model.

We observe that the number of overlaps compared to the number of branch points is rather large (exceeding 10) for certain initial filament orientations. Indeed, the notion of many overlaps does not rule out the notion of branching. A decrease in the capping rate increases $\chi$, as expected, since the filaments are typically longer. Moreover, $\chi$ rather small for the random branch angle as compared to the fixed branch angle model. The lack of inheritance reduces the number of potential overlaps. The reduction in overlaps should also ring true for a completely non-branched Arp2/3 nucleation model as proposed by Urban and collaborators.  In Figure 9, we also plot the overlap for different branch angles. Note that for $\psi=70^{\circ}$, $\chi$ is approximately the same for those filaments originally oriented at $\theta=35^{\circ}$ and for $\theta=90^{\circ}$, the subdominant population of filaments. 

\begin{figure}[b]
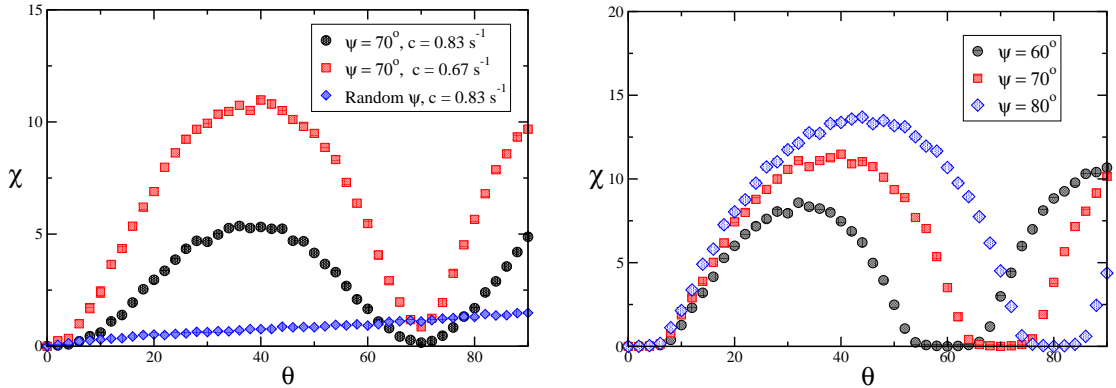

\vspace{0.25cm}
\begin{center}
\includegraphics[width=7cm]{overlap.eps}
\hspace{0.5cm}
\includegraphics[width=7cm]{overlap2.eps}
\caption{Left: The ratio of overlaps to branch points as a function of $\theta$ with $\sigma=10^{\circ}$ for two fixed branch angle curves. Right: The ratio of overlaps to branch points as a function of $\theta$ for a smaller capping rate (more overlaps). Here, $\sigma=0^{\circ}$.}
\end{center}
\end{figure} 

{\bf Filament tip spatial distributions:} Finally, using the two-dimensional discrete simulation, we compute the spatial distribution of filament tips. See Figure 10.  From one filament centered at $x=y=0$, we observe spreading in the $x-$direction of the dendritic array by several microns.  While there is not much difference
between the different branch angles, for the random branch angle model, the broadening in the $x$-direction is enhanced. However, that broadening is not supported by a large number of overlaps making the network more susceptible to
buckling.  As for the $y$-direction, the smaller branch angle allows for more forward growth, as expected, however, the overlap ratio is also smaller. For the random branch angle, the growth in the $y$-direction is the largest, but, again, there is not much structural support via overlaps. 

Combining the overlap data with the distribution of $x$-data we observe that the system is spreading out in the $x$-direction as well as overlapping. The spreading allows for the construction of focal adhesions with which the cells temporarily adhere to the surface. The overlaps enhance structural support. Both features are simultaneously possible in a branched model via the dendritic nucleation model. In the absence of the branches, the system cross-links, albeit not as effectively as a branched model, but does not spread out in the $x$-direction. Moreover, the proliferation of branches ($G$), from a material standpoint, results in the effective strengthening of the material as the meshwork size, the distance between overlaps, decreases with an increasing number of branches. If one were to suspend disassembly of the network, this gradation can be modelled via a spatially varying elastic modulus. More specifically, if one were to model the quasi-two-dimensional lamellipodia as a thin elastic plate with a spatially varying elastic modulus, then (1) the buckling instability softens in that the system does not undergo discontinuous mode changes and (2) the system becomes more robust against out-of-plane buckling as the elastic modulus increases along the direction of axial compression~\cite{Hardik}. Therefore, branching accounts for spreading, reinforcements via overlaps, and gradation.  Only reinforcement is possible in a nonbranching model for a fixed Arp2/3 concentration. 

\begin{figure}[t]
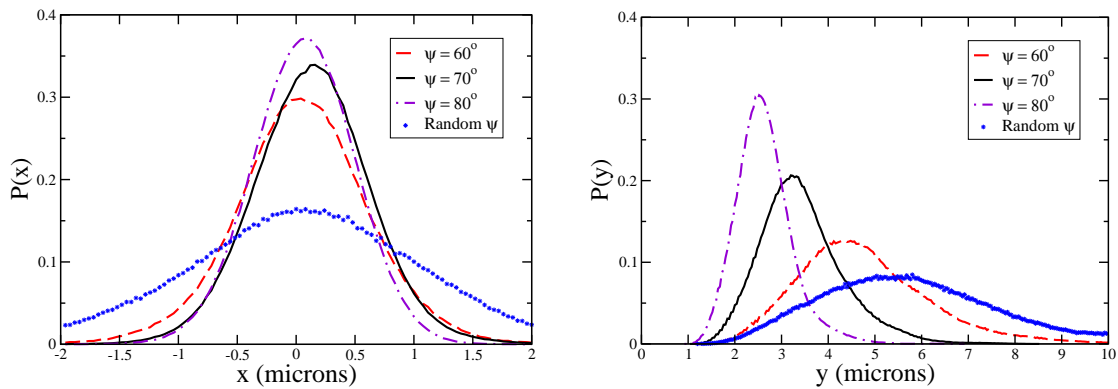

\begin{center}
\includegraphics[width=7cm]{spatial.x.eps}
\hspace{0.5cm}
\includegraphics[width=7cm]{spatial.y.eps}
\caption{Left: Spatial distribution of filament tips in the $x$-direction (horizontal) for $\psi=60^{\circ},70^{\circ},80^{\circ}$ and the random branch model. Right: Spatial distribution of filament tips in the $y$-direction (vertical) for $\psi=60^{\circ},70^{\circ},80^{\circ}$ and the random branch model. For both plots, $\sigma=10^{\circ}$.}
\end{center}
\end{figure} 

\section{Discussion}

Invoking a geometric notion for collision-based branching between globular Arp2/3 and linear, actin filaments, we constructed a mean field model for the orientation of actin filaments near the leading edge of a crawling cell.  To study the model, we applied the approach of Maly and Borisy~\cite{Maly}, who constructed and studied an initial mean field model with a different physical basis than ours. The Maly and Borisy approach~\cite{Maly} invoked a population biology framework with branching corresponding to birth and capping corresponding to death. More specifically, they used the Fisherian criterion~\cite{Fisher,Crow} for maximal reproduction as an optimization condition on the filament orientation. Similar to Maly and Borisy~\cite{Maly}, we found consistency with previous measurements of the distribution of filament orientation with respect to the leading edge. In particular, the two, well-defined peaks in the distribution at $\theta^*=\pm35^{\circ}=\pm\psi/2$ coincide with the optimal relation, assuming $\psi=70^{\circ}$. The fact that both our kinetic model and Maly/Borisy model~\cite{Maly} obtain the same optimal relation despite the differing kinetic assumptions, even in the presence of noise, is interesting and calls for further differentiation. 

Our Arp2/3-actin collision-based model predicts a subdominant population of filaments that may account for recent measurements on the distribution of filament orientation, which are in apparent contradiction with the earlier measurements~\cite{Koestler}. The more recent experiment reported a more broad distribution in filament orientation than previously measured. Moreover, the subdominant population of filaments may be invoked to more accurately model the filament density along the leading edge. Earlier modelling of the filament density demonstrated that larger filament densities in the center required smaller filament densities on the sides of lamellipodia~\cite{Lacayo,Keren}. This requirement is not so consistent with observation, however. By extending the earlier filament density model to include the subdominant population of filaments, this requirement has been relaxed such that the revised filament density model results are more consistent with observations of ``excess'' filament density at the sides of the leading edge. 

To go beyond mean field and study both the positional and orientational degrees of freedom of the actin network in its initial growth phase, we implemented a two-dimensional, kinetic simulation. The mean field optimization condition persists in the two-dimensional simulation, at least for small fluctuations in the branch angle. It would be interesting to extend our collision-based two-dimensional model to include debranching, depolymerization and severing so that we can analyze the approach to steady state and compare our results to the autocatalytic model developed by Carlsson~\cite{Auto}, which was recently verified experimentally~\cite{Kang}. 

Very recent observations of lamellipodia in motile cells via electron tomography reported many more overlaps between filaments than previously estimated using two-dimensional electron micrograph images~\cite{Urban}. Urban and collaborators~\cite{Urban} used this observation to dispute the dendritic nucleation mdoel and propose a new model of unbranched filament nucleation for lamellipodia construction. However, our measurements of the ratio of overlaps to branch points are of order 10 using a branched model.  For a branched model to have no overlaps, the filament lengths must be exponentially decreasing in length with each generation, i.e. the planar embedding of a Bethe lattice in two-dimensions. The rather large ratio of overlaps to branch points actually supports the dendritic nucleation model with its inherited branch angle.  The inheritance increases the potential for overlaps and the pairing of filaments. In fact, the pairing observation was also used by Urban and collaborators~\cite{Urban} as a mark against the dendritic nucleation model. We must also point out that branching also promotes spreading of the network to more readily assemble focal adhesions, and gradation to make the network less susceptible to out-of-plane buckling.  

There exists other evidence for an unbranched model for lamellipodia reconstruction promoting cell motility. Based on experimental observation, Brieher and collaborators~\cite{Brieher} proposed an initial branching motility phase followed by a bundled-actin motility phase, facilitated by facsin or other cross-linking proteins. The notion of a filopodia-dominated phase of motility cannot be ruled out and may be one of many phases of cell motility. However, reconstituted experiments with Arp2/3-actin-fascin demonstrated that Arp2/3 is excluded from the bundling regions~\cite{Vignjevic}. Recent modeling supports this notion~\cite{KCLee}. Unfortunately, Urban and collaborators~\cite{Urban} were unable to determine the spatial location of the Arp2/3 in their experiments. We must also point out that the proposal of unbranched Arp2/3 nucleation implies that the filament density along the leading edge depends purely on the Arp2/3 and not on the pre-existing filament density. If the concentration of Arp2/3 is reasonably uniform along the leading edge, then the filament density profile will also be reasonably uniform near the center of the leading edge. Some such profiles were observed in ``rough'' crawling cells~\cite{Keren}.

Finally, while we have addressed the optimization between the filament orientation and the branch angle, we have not addressed the optimization of the branch angle itself. Why does $\psi\approx 70^{\circ}$?  From the results of this work, only when $\psi>60^{\circ}$ is the redundant second optimum removed, thus paving the way for a suboptimal orientation whose center axis is not $\theta=0^{\circ}$. These off-axis populations allow for further spreading and overlapping. We also observe that the ratio of overlaps to branch points is approximately the same for the optimized orientation of $\theta=\pm 35^{\circ}$ as well as $\theta=\pm90^{\circ}$ so that there is an elastic similarity between the two types of orientations. Other speculations as to why $\psi\approx 70^{\circ}$ may be rooted in structural optimization and the like.

One of us (JMS) would like to acknowledge the hospitality of the Aspen Center of Physics where part of this work was completed and finanical support from DMR-0654373.  Both authors would like to acknowledge helpful discussion with T. Svitkina.

\bibliographystyle{ieeepes}
\nocite{*}
\bibliography{optimal_branching_bibo}

\end{document}